\documentclass{article}

\usepackage{arxiv}

\usepackage[utf8]{inputenc} 
\usepackage[T1]{fontenc}    
\usepackage{hyperref}       
\usepackage{url}            
\usepackage{booktabs}       
\usepackage{amsfonts}       
\usepackage{nicefrac}       
\usepackage{microtype}      
\usepackage{lipsum}
\usepackage{amsmath}
\usepackage{listings}
\usepackage{graphicx}
\usepackage{threeparttablex}
\usepackage{placeins}
\usepackage{mcode}
\usepackage{array}

\title{The Multiplicative Mixed Model with the mumm
R package as a General and Easy Random Interaction
Model Tool}

\author{
  Sofie P\o denphant\thanks{Corresponding author} \\
  Section for Statistics and Data Analysis\\
  DTU Compute \\
  Technical University of Denmark\\
  Kongens Lyngby, Denmark \\
  \texttt{sofp@dtu.dk} \\
   \And
 Kasper Kristensen \\
  Section for Marine Living Resources\\
  DTU Aqua \\
  Technical University of Denmark\\
  Kongens Lyngby, Denmark \\
  \And
   Per B. Brockhoff\\
  DTU Compute\\
  Technical University of Denmark\\
  Kongens Lyngby, Denmark \\
}

\begin{document}
\maketitle

\begin{abstract}
Multiplicative mixed models can be applied in a wide range of scientific disciplines, since they are relevant in every situation where an interaction between a fixed effect and a random effect is present. Until now, no R package has been published, which can fit this type of models. The lack of user-friendly open source tools to fit these models, is the main reason that the models are not used as often as they could or should be. 
In this paper we introduce the user-friendly R package $\mathsf{mumm}$ for fitting multiplicative mixed models in a time-efficient manner. To illustrate the interpretation of the multiplicative term, we provide four data analysis examples, where the model is fitted to data sets that stem from studies in sensometrics, agriculture and medicine. With these examples it is shown that the statistical inference can be improved by using a multiplicative mixed model, instead of a linear mixed model which is usually employed.
\end{abstract}


\section{Introduction}
Linear mixed models are commonly used in a wide variety of disciplines in biological, medical, social, sensory and physical science. They are, however, not always able to describe the complex structure in data sufficiently, and in some cases the data analysis can be improved by extending the linear mixed model with one or more multiplicative term(s), resulting in a so-called multiplicative mixed model \citep{piepho97,piepho99,MAM,  smith01,multSens,smith05}. 
A multiplicative term is in the current work considered to be a product of a random effect and a fixed effect, i.e. a term that models a part of the interaction as a random coefficient model based on linear regression on the fixed effect.

This paper presents three examples on how the multiplicative mixed model can be applied in different fields. The first example is from the field of sensory science, where the objective is to compare certain products, based on scores given to these products by a panel of assessors. For this type of data, the use of the multiplicative mixed model results in a more nuanced modelling and interpretation of the panelist-by-product interaction and improved inference about the product effect. This is possible because the multiplicative term explains the part of the interaction that arises from the panelists using the scoring scale differently. In \citet{MAM} only a linear approximation to the model is fitted, whereas the exact model is fitted in this paper by optimization of the likelihood function. 

The second application is in agriculture, where the multiplicative term in the model accounts for genotypes having different sensitivities to the cultivation environment. The last example in this paper deals with method comparison studies in medicine, where the multiplicative mixed model improves the analysis when the medical methods do not have the same linear calibration. Often in method comparison studies, limits of agreement (LoA) are estimated to assess the agreement between two methods. This paper provides a formula for estimating the LoA under the multiplicative mixed model, together with a simulation study that illustrates the difference between the LoA estimated under the multiplicative mixed model and under the standard mixed two-way ANOVA model.

In general the multiplicative mixed model can be useful in every situation where we have an interaction between
a fixed and a random effect, e.g. in randomized block designs. This implies that there are a lot of applications, where a more correct inference might be possible to achieve. Further, the multiplicative mixed model can be used as a model validation of the assumption of completely unstructured and independent random effects.

However, until now, no open source tools to fit this kind of models exists.
In this paper, the user-friendly R package $\mathsf{mumm}$ \citep{mumm}, for fitting multiplicative mixed models is presented, which finds the maximum likelihood estimated model parameters quickly by making use of R package $\mathsf{TMB}$ (Template Model  Builder) \citep{RTMB}. The syntax of $\mathsf{mumm}$ is very similar to the syntax of the heavily used functions $\mathsf{lm}$ and $\mathsf{lmer}$ \citep[][]{R,Rlme4} for fitting linear and mixed linear models, which makes it straightforward to use.    
\\
\\
The following section gives a detailed formulation of the model and describes how it is estimated by $\mathsf{mumm}$. Section \ref{sec:app} presents the three examples of applications and section \ref{sec:sas} includes a small discussion about computation time, where $\mathsf{mumm}$ and the $\mathsf{NLMIXED}$ procedure in SAS are compared. Finally, concluding remarks are stated in section \ref{sec:conc}.

\section{Method}
The version of the multiplicative mixed model studied in this paper can be written as
\begin{equation} \label{eq:linearMulti}
\begin{gathered} 
\mathbf{y} = \mathbf{X} \boldsymbol\theta + \mathbf{Z}(\boldsymbol\theta) \mathbf{w} + \boldsymbol\epsilon,\\ 
\mathbf{w} \sim \mathcal{N}(\mathbf{0},\mathbf{G}), \ \ \ \boldsymbol\epsilon \sim \mathcal{N}(\mathbf{0},\mathbf{R}),
\end{gathered}
\end{equation} 
where $\boldsymbol\theta$ is a vector containing fixed effects, and  $\mathbf{w}$ is a vector containing the random effects. Moreover $\mathbf{R} =  \sigma^2 \mathbf{I}_n$, where $n$ is the total number of observations.
\\
\\
We will limit our focus to multiplicative models with only one multiplicative term, and only consider models where the fixed effect in the multiplicative term is part of the mean structure, such that $\mathbf{Z}$ contains parameters from $\boldsymbol\theta$, meaning that a fixed effect from the mean structure enters the variance structure. 
\\
In this paper we will mainly use the following two-way model
\begin{equation} \label{eq:MMM}
\begin{gathered}
y_{ijk} = \mu + a_i + \nu_j + b_i \nu_j + d_{ij} + \epsilon_{ijk},\\ i=1,\ldots,I, \ j=1,\ldots,J, \ k=1,\ldots,K,
\\
d_{ij} \sim \ \text{i.i.d.} \ \mathcal{N}(0,\sigma_d^2), \ \ \epsilon_{ijk} \sim \ \text{i.i.d.} \ \mathcal{N}(0,\sigma^2), \\
(a_i,b_i) \sim \mathcal{N} \left(\mathbf{0}, \left[ 
\begin{matrix}
  \sigma_a^2 & \rho \sigma_a \sigma_b \\
  \rho \sigma_a \sigma_b & \sigma_b^2
 \end{matrix}
  \right] \right), 
\end{gathered}
\end{equation}
where we regress on the fixed effect, $\nu_j$, and the regression slopes ($b_i+1$) have mean 1.
\\
\\
In situations where the fixed effect from the multiplicative term is \textit{not} a part of the mean structure, the mean and the variance are not confounded and the covariance structure has a factor-analytic form \citep[][]{piepho97,piepho15}. Such models can be fitted by e.g., the MIXED procedure in SAS, and will not be considered further in this paper.

\subsection{Model Estimation} \label{sec:est}
We want to minimize the negative log-likelihood which is given by
\begin{equation*}
\ell(\boldsymbol\theta,\sigma, \boldsymbol\Psi;\mathbf{y}) = \frac{n}{2}\log(2 \pi) +\frac{1}{2} \log|\mathbf{V}(\boldsymbol\theta) | 
 +\frac{1}{2} (\mathbf{y}-\mathbf{X} \boldsymbol\theta)^T\mathbf{V}(\boldsymbol\theta)^{-1}  (\mathbf{y}-\mathbf{X} \boldsymbol\theta),
\end{equation*}
where $\boldsymbol\Psi$ is a vector containing the variance components of the random effects, and where
\begin{align*}
\mathbf{V}(\boldsymbol\theta) &= Cov(\mathbf{y})   \\ 
&=  \mathbf{Z}(\boldsymbol\theta) \mathbf{G} \mathbf{Z}(\boldsymbol\theta)^T + \mathbf{R}.
\end{align*}
It is important to notice that the covariance matrix contains the fixed effects, since it influences the optimization of the likelihood function.

The standard approach for finding the minimum when working with linear mixed models is to first profile the likelihood so it is a function of the variance components only.
For the multiplicative mixed model, however, the fixed effect parameters cannot be profiled out of the likelihood function, since they, as shown above, are part of the covariance matrix. It is, though, possible to maximize the log-likelihood function iteratively by standard optimization routines, but this might be too time-consuming, depending on the size of the data, due to the inversion of the covariance matrix.

Alternatively, the multiplicative mixed model can be estimated by maximizing the Laplace approximation to the log-likelihood, which is advantageous regarding computation speed. The Laplace approximation is exact for this model, due to the data being Gaussian, which means that we can gain speed-ups without loosing accuracy, and we will therefore pursuit this approach.
The Laplace approximation can be written as:
\begin{equation*}
\ell_{LA}(\boldsymbol\theta,\sigma, \boldsymbol\Psi;\mathbf{y}) = h(\boldsymbol\theta, \sigma, \boldsymbol\Psi; \tilde{\mathbf{w}}, \mathbf{y}) - \frac{1}{2} \log\left( \left|\frac{ -\mathbf{H} (\boldsymbol\theta, \sigma , \boldsymbol\Psi, \tilde{\mathbf{w}}) }{2 \pi}\right|\right),
\end{equation*}
where $h(\boldsymbol\theta, \sigma, \boldsymbol\Psi; \mathbf{w}, \mathbf{y})$ is the joint log-likelihood function, $\mathbf{H} (\boldsymbol\theta, \sigma , \boldsymbol\Psi, \mathbf{w})$ is the Hessian of the joint log-likelihood function with respect to $\mathbf{w}$, and $\tilde{\mathbf{w}}$ is the maximizer of $h$.
\\
The optimization of $\ell_{LA}$ for the multiplicative mixed model can be done by making use of the R package $\mathsf{mumm}$.

\subsection{R package $\mathsf{mumm}$}
The R package $\mathsf{mumm}$ \citep{mumm} makes it possible to fit multiplicative mixed models. The package provides a function where the user only needs to give a model formula and the data set as input to get the estimated model fit and standard model summaries as output. The syntax for the model formula is very similar to the syntax in the $\mathsf{lmer}$-function for fitting linear mixed models by the $\mathsf{lme4}$ R package, which makes package $\mathsf{mumm}$ very user-friendly. For the optimization part, the package makes use of the $\mathsf{TMB}$ package, which will be described in the following section. 

\subsubsection{The Template Model Builder ($\mathsf{TMB}$)} 
The Template Model Builder ($\mathsf{TMB}$) is a recently developed R package \citep{RTMB} that enables fast maximization of the Laplace approximation to the marginal log-likelihood function of nonlinear mixed models. The user needs to define the negative joint log-likelihood function in a C++-template, while the remaining code is written in R. The package uses Automatic Differentiation (AD) to compute the derivatives of the joint log-likelihood function with respect to the random effects, which are used to build the Laplace approximation and its gradient. First order derivatives are usually sufficient for the maximization of likelihood functions, but since the Laplace approximation involves up to second order derivatives, up to third order derivatives are necessary to compute its gradient. This is facilitated by TMB by clever application of automatic differentiation.
The Laplace approximation and its gradient are then given as input to a minimizer in R, e.g. the R-function $\mathsf{nlminb}$ \citep{R}, which optimizes the Laplace likelihood and returns the maximum likelihood estimates of the parameters.

By using R package $\mathsf{mumm}$ to fit the multiplicative mixed model, the user exploits the speed of $\mathsf{TMB}$ but avoids the coding of C++-templates.

\section{Applications} \label{sec:app}
In this section, we provide four data analysis examples, where the $\mathsf{mumm}$ package is used to fit the multiplicative mixed model to data from sensometrics, agriculture and medicine.

\subsection{Sensory Science}
In the field of sensory science, the use of models with multiplicative terms has been proposed and discussed in several publications \citep[see, e.g.,][]{MAM,per94,multSens}. Sensory profile data, where $I$ assessors scored $J$ products in $K$ replicates, is frequently analysed by a linear mixed two-way ANOVA model.
The model contains the overall mean, the random effect of assessor, a fixed product effect, a random panelist-by-product disagreement effect and the random residual error. It has been widely discussed in the sensory literature whether to consider the assessor effect as random or fixed, but as argued in \citet{MAM}, the usual purpose of the experiment is to draw inference about the tested products that may generalize to a larger setting than the assessors entering the panel, and a common approach is thus to consider the effects related to the assessors as random.
\\
\\
The panelist-by-product interaction models the differences in perception of the products but also the differences in the assessors individual ranges of scale use. To account for these individual ranges of scale use, \citet{per94} introduces a multiplicative term as an extra term in the two-way ANOVA model without interaction. The assessors are in this approach considered to be a fixed effect, which simplifies the estimation of the model. In \citet{MAM} however, the model is introduced in an extended version where the standard interaction is included and where the assessor dependent effects are considered to be random. The purpose of this model is to partition the panelist-by-product interaction into an assessor dependent scaling effect and the actual disagreement effect that is due to the differences in perception of the products. The formal multiplicative mixed model, which we are fitting, is not fitted in \citet{MAM}. Instead they suggest a linear approximate approach based on the so called Mixed Assessor Model (MAM):
\begin{equation} \label{eq:MAM}
\begin{gathered} 
y_{ijk} = \mu + a_i + \nu_j + \beta_i x_j + d_{ij} + \epsilon_{ijk}, \\
a_i \sim \text{i.i.d.} \ \mathcal{N}(0,\sigma_a^2), \ \  d_{ij} \sim \text{i.i.d.} \ \mathcal{N}(0,\sigma_d^2), \\ \epsilon_{ijk} \sim \text{i.i.d.} \ \mathcal{N}(0,\sigma^2),
\end{gathered}
\end{equation}
where $x_j = \overline{y}_{.j.}-\overline{y}_{...}$ is the centered product averages for product $j$, included as a covariate, and the regression coefficient, $\beta_i$, is the individual scaling slope for assessor $i$, with $\sum_{i = 1}^I\beta_i = 0$.
The MAM is a simple linear mixed model since the $x_j$s are found directly from data before estimating the model and are therefore considered known. This is justified by the fact that this model still produces valid hypothesis tests for product differences. However, this model will in general not produce valid product difference confidence intervals, since the $\beta$s are considered fixed instead of random, even though they are assessor dependent.
\\
\\
The formal multiplicative mixed model can be written as (\ref{eq:MMM}),
where $a_i$ is the main effect of assessor, $\nu_j$ is the product effect, $b_i$ is the assessor specific scaling effect and $d_{ij}$ is the panelist-by-product disagreement effect. By including the multiplicative term, we perform a regression for each assessor against the centered product effects - the larger the range of scale use, the larger the regression coefficient ($1+b_i$). It is assumed that observations across assessors are independent and that $\epsilon_{ijk}$ is independent of the random effects. Under this model, profile likelihood based confidence intervals for the product differences can easily be obtained with R package $\mathsf{mumm}$, together with the model parameter estimates. 

\subsubsection{Data Example: B\&O TV Data} \label{sec:BO}
The example data set is acquired from the $\mathsf{lmerTest}$-package in R \citep[][]{RlmerTest,RlmerTestA} and it stems from a sensory evaluation of a series of Bang \& Olufsen televisions. The televisions are assessed based on 15 characteristics such as \textit{colour balance, colour saturation, noise, sharpness, cutting} etc., but in this example we restrict ourselves to one single characteristic - namely \textit{cutting}. The term cutting refers to how much of the picture that has been cut off on the screen, and the scale goes from nothing to much.

The purpose is to test the products, which are specified by two attributes \textit{TVset} and \textit{Picture} having three and four levels respectively. These attributes are crossed, which gives us one combined product factor with 12 levels (J = 12).
\\
\\
The 12 products were assessed by a panel of eight assessors (I = 8), scoring each product in two replicates (K = 2) - yielding a total of 192 measurements. It is assumed that the replications are completely randomized, hence we assume that no block effect is present.
\\
\\
Using R package $\mathsf{mumm}$, the multiplicative mixed model is fitted to the data by the following code, where $\mathsf{mp(Assessor,Product)}$ is the syntax for the multiplicative term.

\begin{figure}[h!]
\centering
  \includegraphics[width=0.75\textwidth]{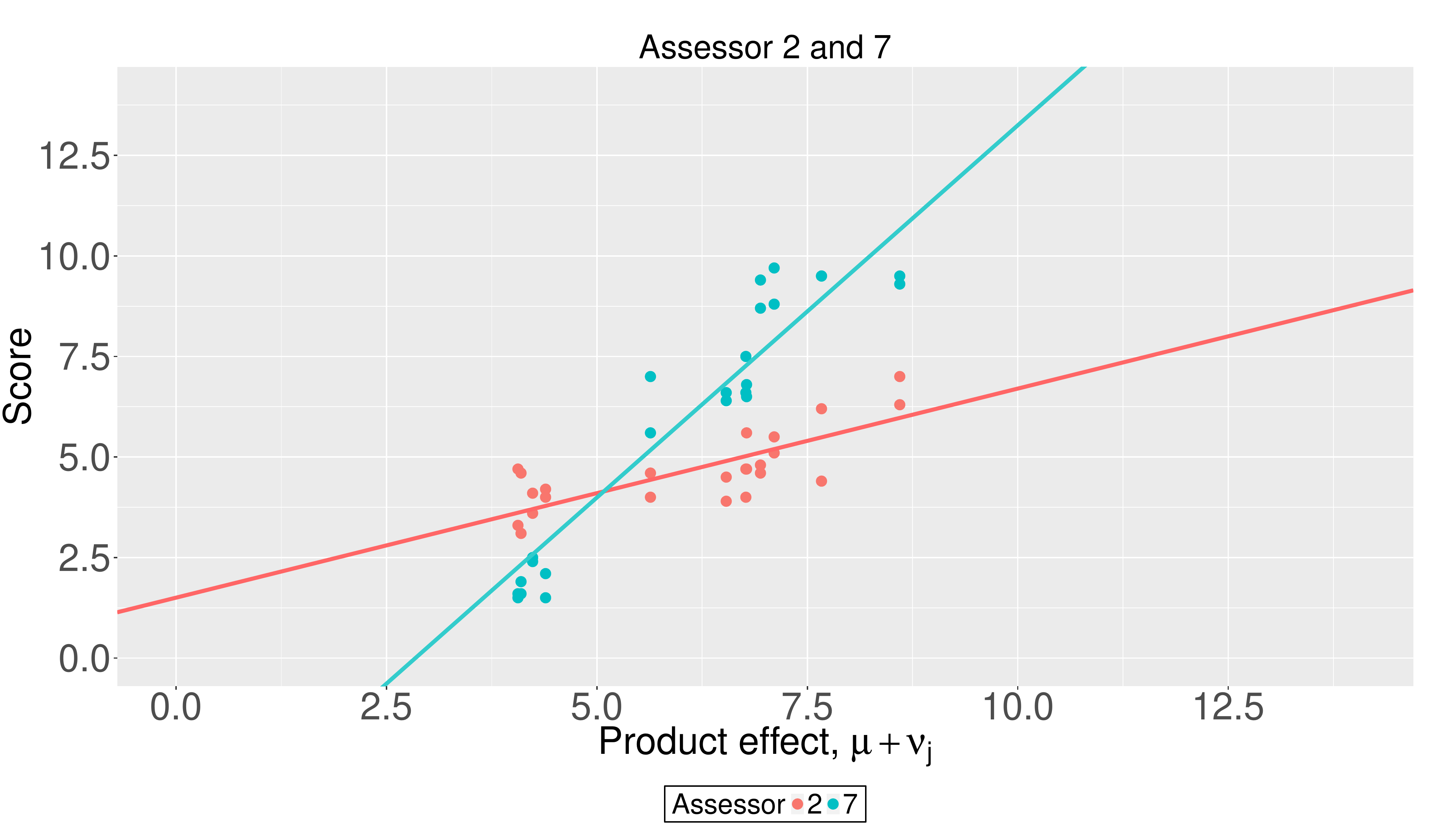}
  \caption{The data points and the fitted regression lines for Assessor 2 and 7 in the B\&O TV data set.}\label{fig:fitted_2_7}
 \end{figure}

\begin{figure}[htb!]
  \centering
  \includegraphics[width=0.75\linewidth]{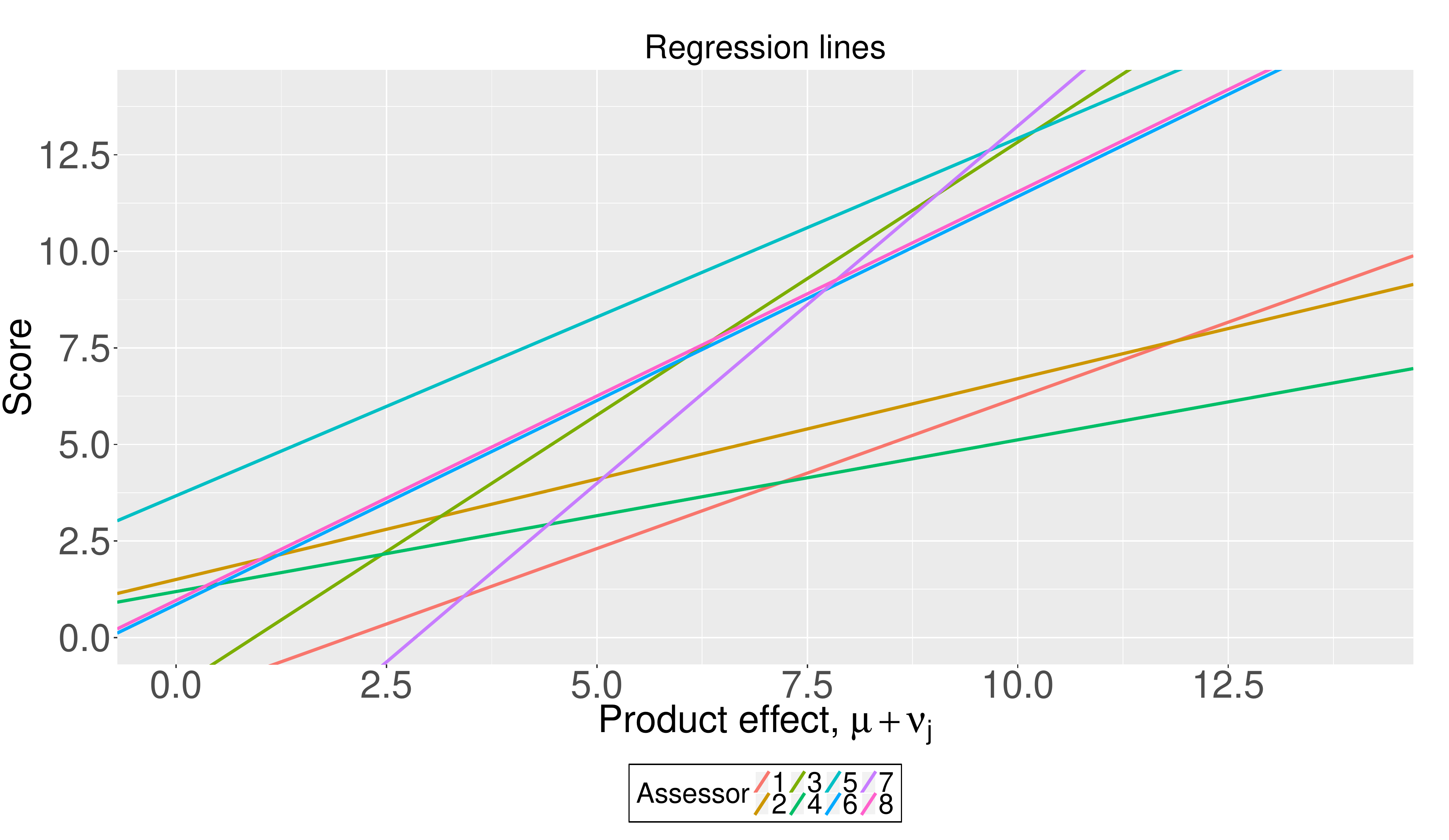}
  \caption{The fitted regression lines for the B\&O TV data set. The slopes of the lines are $b_i + 1$ and the intercepts are $a_i - \mu b_i$.}\label{fig:fitted_all}
\end{figure}

\begin{lstlisting}[numbers = none, language = R, basicstyle = \ttfamily\small]
fit = mumm(Cutting ~ 1 + Product + (1|Assessor) + 
			(1|Assessor:Product) + mp(Assessor,Product), data = BO)
\end{lstlisting}
Appendix \ref{app} contains the full code used to analyse this data set.
\begin{table}[htb!]
\caption{\small{The estimated model parameters and the random effect estimates.}} \label{tab:parTV}
\small
\begin{center}
\begin{tabular}{cccccccccccc}
\hline\noalign{\smallskip}
$\mu+\nu_1$ & $\mu+\nu_2$ & $\mu+\nu_3$ & $\mu+\nu_4$ & $\mu+\nu_5$ & $\mu+\nu_6$ \\ 
\noalign{\smallskip}\hline\noalign{\smallskip}
7.1057 & 8.5980 & 7.6681 & 6.9428 & 6.5361 & 6.7693  \\
\noalign{\smallskip}\hline
\hline\noalign{\smallskip}
$\mu+\nu_7$ & $\mu+\nu_8$ & $\mu+\nu_9$ & $\mu+\nu_{10}$ & $\mu+\nu_{11}$ & $\mu+\nu_{12}$  \\ 
\noalign{\smallskip}\hline\noalign{\smallskip}
5.6357 & 6.7778 & 4.3898 & 4.2358 & 4.0981 & 4.0615 \\ 
\noalign{\smallskip}\hline
\end{tabular}
\\[10pt]
\begin{tabular}{cccccccc}
\hline\noalign{\smallskip}
$a_1$ & $a_2$ & $a_3$ & $a_4$ & $a_5$ & $a_6$ & $a_7$ & $a_8$  \\ 
\noalign{\smallskip}\hline\noalign{\smallskip}
-2.9305 & -1.4115 & 1.2018 & -2.4946 & 3.2196 & 1.1976 & -0.0986 & 1.3161\\
\noalign{\smallskip}\hline
\end{tabular}
\\[10pt]
\begin{tabular}{cccccccc}
\hline\noalign{\smallskip}
$b_1$ & $b_2$ & $b_3$ & $b_4$ & $b_5$ & $b_6$ & $b_7$ & $b_8$\\ 
\noalign{\smallskip}\hline\noalign{\smallskip}
-0.2183 & -0.4802 & 0.4141 & -0.6071 & -0.0742 & 0.0569 & 0.8508 & 0.0581 \\
\noalign{\smallskip}\hline
\end{tabular}
\\[10pt]
\begin{tabular}{ccccc}
\hline\noalign{\smallskip}
$\sigma$ & $\sigma_a$ & $\sigma_b$ & $\sigma_d$ & $\rho$ \\
\noalign{\smallskip}\hline\noalign{\smallskip}
1.2100 & 2.0096 & 0.4692 & 0.2428 & 0.4188\\ 
\noalign{\smallskip}\hline
\end{tabular}
\end{center}
\end{table}
\noindent
\\
The parameter estimates are shown in Table \ref{tab:parTV}, where it is seen that assessor 7 have the largest scaling coefficient ($b_7$) and that assessor 2 has the second most negative scaling coefficient. These coefficients can be interpreted as the deviation from the consensus slope equal to one. The fitted regression lines for these two assessors versus the consensus product pattern are plotted together with the corresponding data points in Figure \ref{fig:fitted_2_7}. The slopes of the lines are $b_i + 1$ and the intercepts are $a_i - \mu b_i$. These lines clearly illustrate the scale range differences between the two assessors. Figure \ref{fig:fitted_all} shows a plot of the fitted regression lines for all of the eight assessors, where the difference in the slope of the lines is very clear, which justifies the use of the multiplicative model.
\noindent
\\
\\
Table \ref{tab:qhi} shows the results of likelihood ratio tests of the significance of the models terms.
Following \citet{self} the test statistic, when testing the significance of a variance parameter, follows an equal mixture of a $\chi^2_0$- and $\chi^2_1$-distribution, approximately. Therefore the degrees of freedom in the test is chosen to be 1/2. Similarly, when testing the significance of a variance parameter and a covariance parameter at the same time, the degrees of freedom in the test is chosen to be 3/2, since \citet{stram} argues that the test statistic in such a test approximately follows an equal mixture of a $\chi^2_1$- and $\chi^2_2$-distribution. 
Not surprisingly, it is seen that the scaling effect \textit{is} significant. The disagreement effect is not significant, meaning that there is no significant difference in the assessors perception of the products; the entire panelist-by-product interaction can be explained by the assessors individual ranges of scale use.
\begin{table*}[t]
\caption{Likelihood Ratio Test of the significance of the model terms. $M_0$ and $M_1$ are the null model and the alternative model, respectively.} \label{tab:qhi}
\small
\begin{center}
\begin{tabular}{lllrrr}
\hline\noalign{\smallskip}
Effect & $M_0$ & $M_1$ & $\chi^2$ & DF & p-value \\
\noalign{\smallskip}\hline\noalign{\smallskip}
Disagreement, $d$ & (\ref{eq:MMM}) without $d$  & (\ref{eq:MMM}) & 0.13 & 1/2 & $4.49E-01$ \\ 
Scaling, $b$ & 2-way ANOVA & (\ref{eq:MMM}) & 28.23 & 3/2 & $3.07E-07$ \\ 
Assessor, $a$ & (\ref{eq:MMM}) without $a$  & (\ref{eq:MMM}) & 133.77 & 3/2 & $0.00E+00$ \\ 
Product, $\nu$ & (\ref{eq:MMM}) without $v$ and $b$ & (\ref{eq:MMM}) & 120.29 & 12.5 & $0.00E+00$ \\ 
\noalign{\smallskip}\hline
\end{tabular}
\end{center}
\end{table*}

\paragraph{Product Differences} 
\noindent
\\
The variance of the product contrast in the multiplicative mixed model given by (\ref{eq:MMM}) is
\begin{equation} \label{eq:MMM_var}
\text{var}(\overline{y}_{.1.}-\overline{y}_{.2.}) = \frac{1}{I}\sigma_b^2 (\nu_1-\nu_2)^2 + \frac{2}{I}\sigma_d^2 + \frac{2}{KI} \sigma^2,
\end{equation}
whereas the variance in MAM given by (\ref{eq:MAM}) is
\begin{equation} \label{eq:MAM_var}
\text{var}(\overline{y}_{.1.}-\overline{y}_{.2.}) = \frac{2}{I}\sigma_d^2 + \frac{2}{KI} \sigma^2
\end{equation}
\citep{MAM}
\\
We see that (\ref{eq:MMM_var}) has an extra term that makes the variance of the product difference depend on the actual size of the difference itself. Under the null hypothesis of no product difference, however, the term disappears and the variance expressions for the two models are equal. This means that the null-hypothesis test under the MAM is also valid under the multiplicative mixed model.      

In (\ref{eq:MAM_var}) the main part of error stems from the disagreement variation, and not from the interaction, which is the case in a mixed two-way ANOVA model. Therefore the F-statistic
\begin{equation} \label{eq:F}
F_{Product} = \frac{MS_{Product}}{MS_{Disagreement}},
\end{equation}  
will follow an F-distribution with $(J-1,(I-1)(J-2))$ degrees of freedom under the null hypothesis and will be a valid test for product difference under the multiplicative mixed model \citep{MAM}. 

Note that the denominator in the F-statistic is $MS_{Disagreement}$ and not $MS_{Interaction}$ as in the mixed two-way ANOVA model, which means that the error used for making inference about product differences in (\ref{eq:F}) is now cleaned out for potential scaling structure. This results in greater power for detecting product differences compared to the standard mixed two-way model, which will be illustrated in the following example. 

To see the actual effect of the increased power, the same data set as before will be used but this time with the characteristic \textit{Sharpness of movement} as the response variable. The term \textit{Sharpness of movement} refers to how sharp the picture is during movement or panning and the scale goes from sharp to unsharp. 

Table \ref{tab:power} shows the computed p-values for testing the significance of the product effect under the three different models. In the likelihood ratio test the null model is the same as in Table \ref{tab:qhi}, i.e.
\begin{gather*}
\text{M}_0: \ \ y_{ijk} = \mu + a_i + d_{ij} + \epsilon_{ijk}, \\ \nonumber
a_i \sim \text{i.i.d.} \ \mathcal{N}(0,\sigma_a^2), \ \  d_{ij} \sim \text{i.i.d.} \ \mathcal{N}(0,\sigma_d^2), \\ \epsilon_{ijk} \sim \text{i.i.d.} \ \mathcal{N}(0,\sigma^2),
\end{gather*}
and the alternative model is the full multiplicative mixed model (\ref{eq:MMM}). 
\\
\\
It is seen that both the likelihood ratio test and the F-test under the MAM result in a lower p-value compared to the F-test under the mixed two-way model.  
\begin{table}[t]
\caption{Computed p-values for testing the significance of an overall product difference with \textit{Sharpness of movement} as the response variable.} \label{tab:power}
\small
\begin{center}
\begin{threeparttable}
\begin{tabular}{lllccccccccc}
\hline\noalign{\smallskip}
Method & Model & Effect & F & $\chi^2$ & DF & p-value \\ 
\noalign{\smallskip}\hline\noalign{\smallskip}
F-test & 2-way ANOVA & Product & 3.74 & - & 11 & $2.67E-4$\\
F-test\tnote{a} & MAM (\ref{eq:MAM}) & Product & 3.86 & - & 11 & $2.28E-4$\\ 
LRT & MMM (\ref{eq:MMM}) & Product & - & 38.60 & 12.5 & $1.70E-4$\\
\noalign{\smallskip}\hline
\end{tabular}
 \begin{tablenotes}
 	\item[a] The results are found by R package $\mathsf{SensMixed}$  \citep{RSensMixed}. 
  \end{tablenotes}
\end{threeparttable}
\end{center}
\end{table}

\subsection{Genotype-by-environment Data}
Genotype-by-environment data is frequently analysed by a two-way ANOVA model \citep[see, e.g.,][]{AgriMalosetti,AgriAus,piepho97,perkins68},
which contains the overall mean of the phenotypic response, the effect of the genotype, the effect of the environment, the genotype-by-environment interaction and the random residual error.
\\
\\
It might though result in a more correct inference to describe the interaction more detailed, and take into consideration that some genotypes are more sensitive to the environment than others.
In \citet{piepho97} a multiplicative model is proposed for the analysis of this type of data to account for this and in \citet{AgriMalosetti} it is argued that genotypes, depending on the aim of the analysis, can be regarded as a random sample from a larger population. 

This leads us to the multiplicative mixed model in (\ref{eq:MMM}), where a part of the genotype-by-environment interaction is modelled by a genotype-specific linear regression on the environmental effects to account for the varying sensitivities of the different genotypes. In this setting, $a_i$ and $b_i$ is the main effect and the sensitivity of genotype $i$, respectively. Further, $\nu_j$ is the fixed main effect of environment $j$ and $d_{ij}$ is the part of the interaction that is not explained by the linear regression. 

\subsubsection{Data Example: The Height of Wheat}
\begin{figure}[htb!]
\centering
  \includegraphics[width=0.75\linewidth]{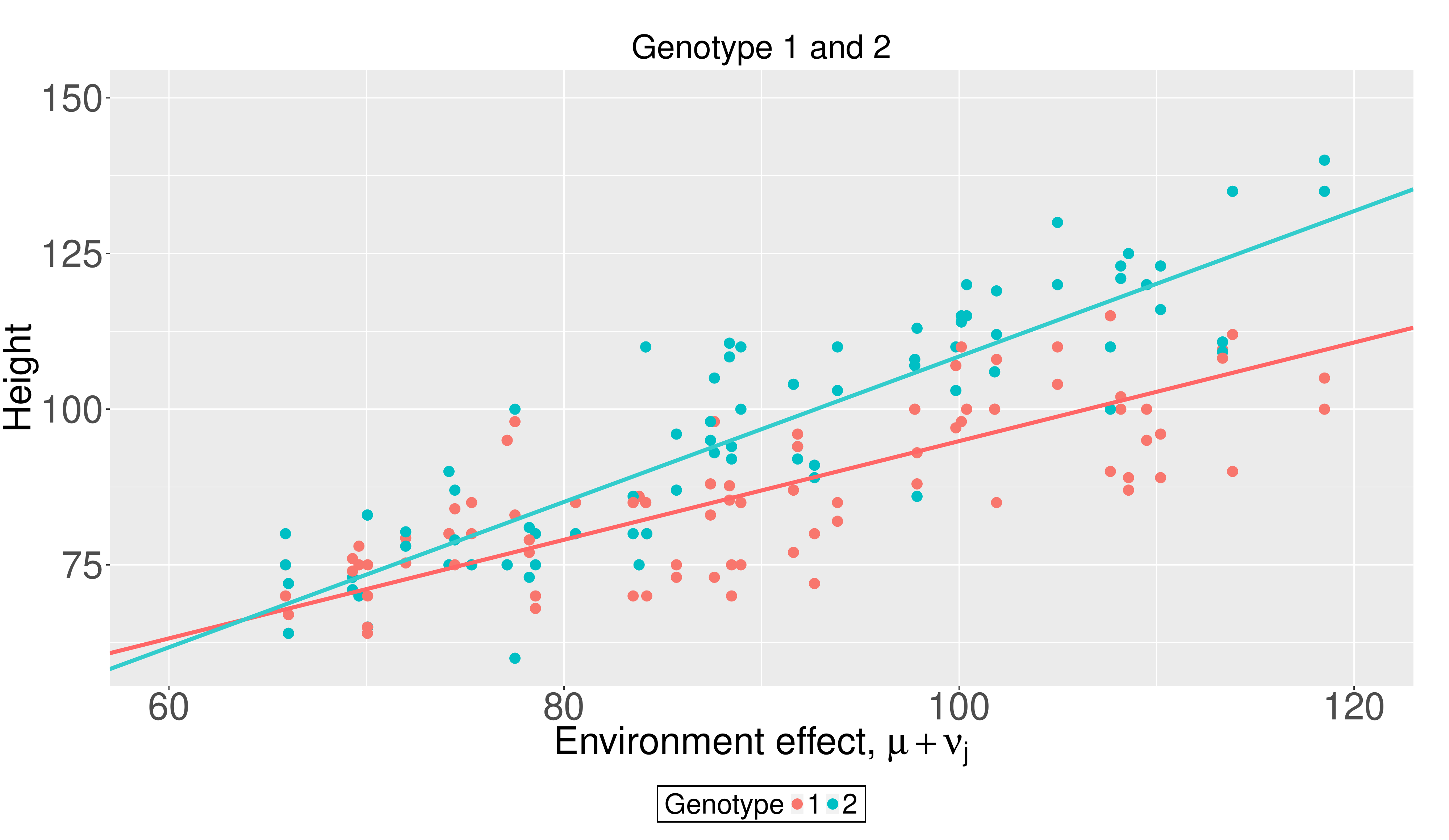}
   \caption{The fitted regression lines and the data points for genotype 1 and 2. The slopes of the lines are $b_i + 1$ and the intercepts are $a_i - \mu b_i$.} \label{fig:scaleWheat}
\end{figure}

A data set from the international maize and wheat improvement center \citep{CIMMYT} will now be analysed by the multiplicative mixed model in (\ref{eq:MMM}). This data set contains the measured height of 50 wheat plants with unique genotypes grown in 45 different environments and there are two replicates. Hence, the phenotypic response in the model is in this case the height of the plant.  
\\
\\
The multiplicative mixed model is fitted to the data and the variance component estimates are shown in Table \ref{tab:parWheat}. The data points for two of the 50 genotypes are plotted together with the corresponding fitted regression lines in Figure \ref{fig:scaleWheat}, where it is seems evident that the two wheat genotypes have different sensitivities to the environment.

\begin{table}[htb!]
\caption{\small{The estimated covariance parameters.}} \label{tab:parWheat}
\small
\begin{center}
\begin{tabular}{ccccc}
\hline\noalign{\smallskip}
$\sigma$ & $\sigma_a$ & $\sigma_b$ & $\sigma_d$ & $\rho$ \\
\noalign{\smallskip}\hline\noalign{\smallskip}
5.1958 & 2.4242 & 0.0691 & 3.1660 & 0.5844  \\ 
\noalign{\smallskip}\hline
\end{tabular}
\end{center}
\end{table}

\noindent
Table \ref{tab:qhiWheat} shows the results of likelihood ratio tests of the significance of the terms, where it seen that all of the model terms are significant, including the the sensitivity effect.

\begin{table}[ht]
\caption{Likelihood Ratio Test of the significance of the model terms.} \label{tab:qhiWheat}
\small
\begin{center}
\begin{tabular}{lrrr}
\hline\noalign{\smallskip}
Effect & $\chi^2$ & DF & p-value \\ 
\noalign{\smallskip}\hline\noalign{\smallskip}
Unexplained interaction, $d$ & 168.10 &  1/2 & $0.00E+00$ \\ 
Sensitivity, $b$  & 50.86 & 3/2 & $3.25E-12$ \\ 
Genotype, $a$ & 386.82 & 3/2 & $0.00E+00$  \\ 
Environment, $\nu$ & 5039.37 & 45.5 & $0.00E+00$   \\ 
\noalign{\smallskip}\hline
\end{tabular}
\end{center}
\end{table}

\subsection{Method Comparison Studies - agreement between random methods}
\noindent
Method comparison studies are studies that are designed to compare different medical instruments or different methods of clinical measurement.
The usual approach when comparing multiple measurement methods is to apply the additive two-way ANOVA model \citep{hawkins2010}:
\begin{gather} \label{2way}
y_{ij} = \mu_j + \alpha_i + \epsilon_{ij}, \  \
\epsilon_{ij} \sim \ \text{i.i.d.} \ \mathcal{N}(0,\sigma^2),
\end{gather}    
where $y_{ij}$ is the measured value obtained from using method $i$ on item $j$, $\mu_j$ is the true value for item $j$ and $\alpha_i$ is the relative bias related to method $i$.                                           
This is also the underlying model in the so-called Bland-Altman setup, which is often used when only two measurement methods are compared and only one measurement by each method is carried out on the items \citep{carstensenBio}. However, if replicate measurements are available, the model can easily be expanded with an interaction term that separates the item-method interaction from the error term. In the rest of this section we will, however, only consider data sets without replicates.
\\ \\
%
Model (\ref{2way}) is not capable of handling situations where the methods do not have the same linear calibration. In such cases, parts of the interaction can be modelled by regressing on the item values \citep[][]{carstensenBio, hawkins2010}. The model is therefore expanded to:
\begin{gather*} 
y_{ij} = \mu_j + \alpha_i + \beta_i \mu_j + \epsilon_{ij}, \  \
\epsilon_{ij} \sim \ \text{i.i.d.} \ \mathcal{N}(0,\sigma^2),
\end{gather*}   
and
where the $\beta_i$'s are the method specific regression coefficients. 
\\
\\
In some experiments, the used measurement methods can be considered a random sample from a larger population of possible measurement methods. In \citet{ekstrom} such a situation is considered. In that case, it is more reasonable to consider the method-specific effects as random and thus to fit the following multiplicative mixed model:
\begin{gather*}
y_{ij} = \mu_j + \tilde{a}_i + b_i \mu_j + \epsilon_{ij}, \  \
\epsilon_{ij} \sim \ \text{i.i.d.} \ \mathcal{N}(0,\sigma^2), 
\\[6pt]
(\tilde{a}_i,b_i) \sim \mathcal{N} \left(\mathbf{0}, \left[ \begin{matrix}
  \sigma_{\tilde{a}}^2 & \tilde{\rho} \sigma_{\tilde{a}} \sigma_b \\
  \tilde{\rho} \sigma_{\tilde{a}} \sigma_b & \sigma_b^2
 \end{matrix}
  \right] \right), \nonumber
\end{gather*}   
where the method dependent parameters are now written in Latin letters to clarify that they are considered random.  
\\
\\
This multiplicative model is written without a term accounting for the overall mean, but when we allow the method specific random effects, $a_i, b_i$, to correlate, this model is similar to (\ref{eq:MMM}), if we discard $d_{ij}$ and index $k$, due to the lack of replicate measurements. In this setting, $\mu_j = \mu+\nu_j$ and $\tilde{a}_i = a_i - b_i \mu$.

\subsubsection{Data Examples}
\textbf{Example 1} \\
In the first data set (kindly supplied by Prof. Douglas Hawkins) \citep{hawkins2010}, six methods for measuring the concentration of hepatitis virus in blood samples are compared. The number of blood samples (items) in the experiment is 51. The experiment is not replicated and the data is incomplete, resulting in only 248 observations instead of 306. 

\begin{figure}[htb]
\centering
  \includegraphics[width=0.75\linewidth]{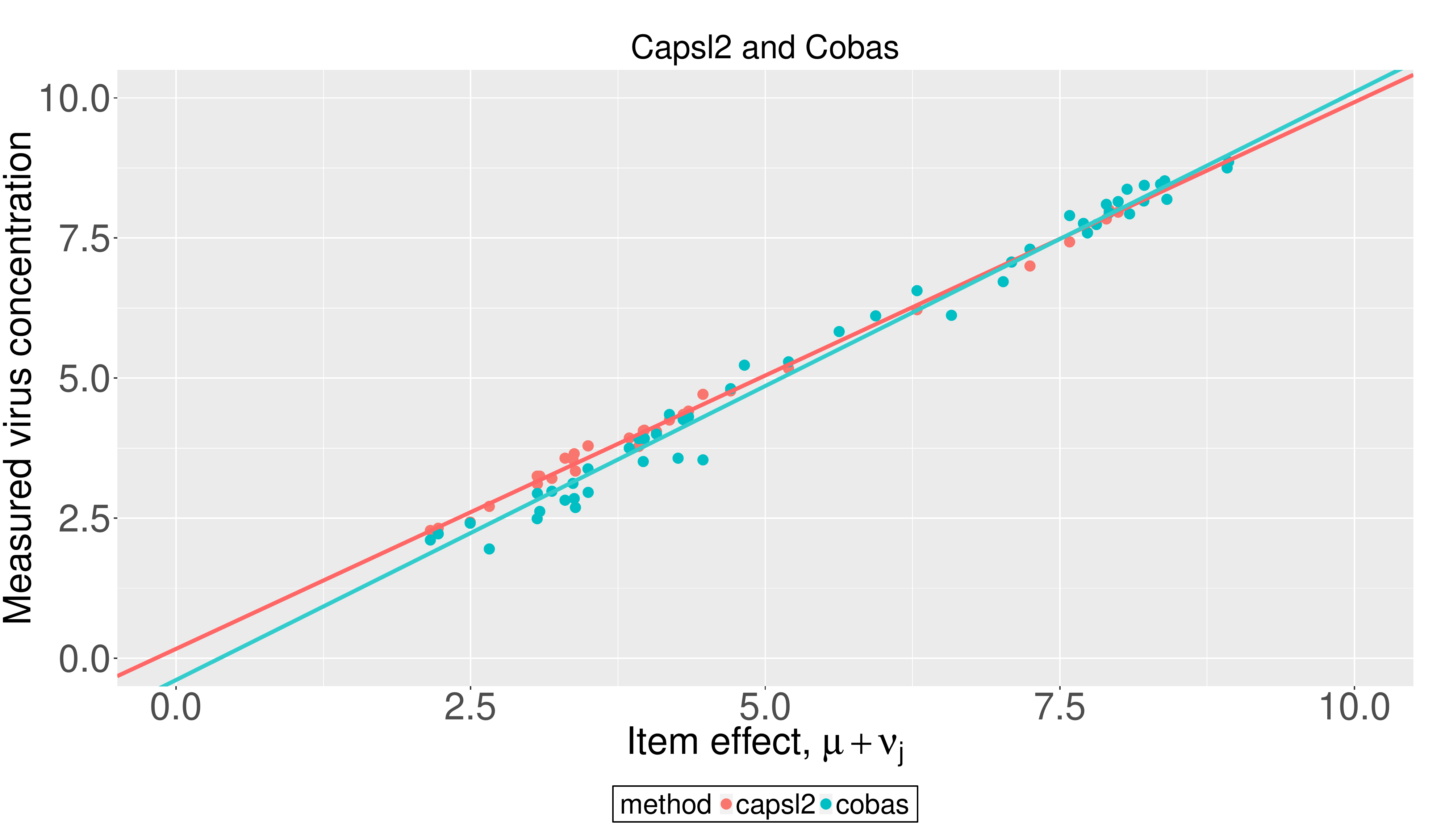}
   \caption{The fitted regression lines and the data points for method Capsl2 and Cobas. The slopes of the lines are $b_i + 1$ and the intercepts are $a_i - \mu b_i$.}\label{fig:scale_HBV}
\end{figure}

The multiplicative mixed model in (\ref{eq:MMM}) is fitted to the data, and Table \ref{tab:var_HBV} shows the estimated variance components in the model and Table \ref{tab:ran_HBV} shows the estimated method specific random coefficients.
\begin{table}[hb]
\caption{The estimated covariance parameters.} \label{tab:var_HBV}
\small
\begin{center}
\begin{tabular}{cccc}
\hline\noalign{\smallskip}
$\sigma$ & $\sigma_a$ & $\sigma_b$ & $\rho$ \\ 
\noalign{\smallskip}\hline\noalign{\smallskip}
 0.2092 & 0.1248 & 0.0275 & -0.5349 \\ 
\noalign{\smallskip}\hline
\end{tabular}
\end{center}
\end{table}
\begin{table}[hb]
\caption{The estimated random coeffiecients.} \label{tab:ran_HBV}
\begin{center}
\begin{tabular}{crrrrrr}
\hline\noalign{\smallskip}
$i$ & capsl1 & capsl2 & capsld & cobas  &  hps & versant  \\ 
\noalign{\smallskip}\hline\noalign{\smallskip}
$a_i$  & -0.1107 & 0.0358 & -0.1044 & -0.1220 & 0.1146 & 0.1868  \\
$b_i$ & 0.0047 & -0.0244 & -0.0018 & 0.0492 & -0.0158 & -0.0120  \\ 
\noalign{\smallskip}\hline
\end{tabular}
\end{center}
\end{table}
\\
We see that the \textit{cobas} method has the steepest regression line and \textit{capsl2} the flattest. Figure \ref{fig:scale_HBV} shows the fitted regression lines for these two methods.
\noindent
Table \ref{tab:qhiHBV} shows the likelihood ratio tests of the significance of the model terms, where it is seen that everything is significant.
\begin{table}[ht]
\caption{Likelihood Ratio Test of the significance of the model terms.} \label{tab:qhiHBV}
\small
\begin{center}
\begin{tabular}{lrrr}
\hline\noalign{\smallskip}
Effect & $\chi^2$ & DF & p-value \\ 
\noalign{\smallskip}\hline\noalign{\smallskip}
Scaling, $b$  & 11.01 & 3/2 & $2.08E-03$ \\ 
Method, $a$ & 62.01 & 3/2 & $1.18E-14$   \\ 
Item, $\nu$ & 1100.04 & 51.5 & $0.00E+00$ \\ 
\noalign{\smallskip}\hline
\end{tabular}
\end{center}
\end{table}
\\
\\
\noindent
\textbf{Example 2}
\\
The second dataset $\mathsf{glucose}$ is from the R package $\mathsf{MethComp}$ \citep{Rmethcomp}. The data is unbalanced and consists of data from 74 persons (items) and four method types. Each person had blood sampled at 0, 30, 60 and 120 min after a 75 g glucose load and the glucose concentration was measured, but we only consider the data sampled after 120 min, with 328 observations.

\begin{figure}[htb!]
\small
\centering
  \includegraphics[width=0.75\linewidth]{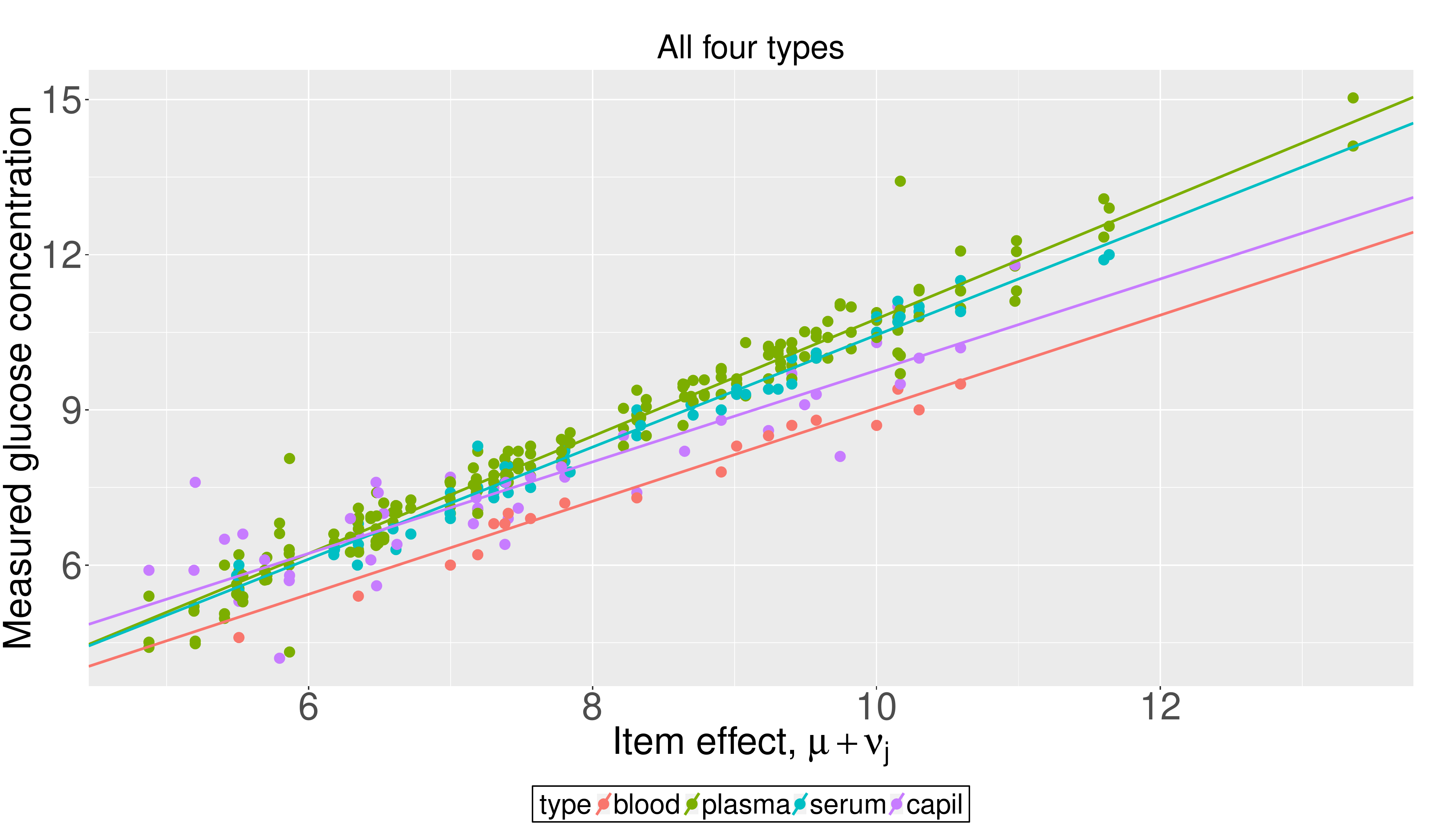}
   \caption{The fitted regression lines and the data points for the glucose data set. The slopes of the lines are $b_i + 1$ and the intercepts are $a_i - \mu b_i$}\label{fig:scale_glucose}
\end{figure}

Table \ref{tab:var_glucose} shows the estimated variance components for the multiplicative mixed model fit where the correlation between the random effects seems to be quite strong. Table \ref{tab:ran_glucose} shows the estimated method specific random parameters, where it is seen that the slope differs the most between \textit{plasma} and \textit{capil}. This difference is quite clear in Figure \ref{fig:scale_glucose}, showing the data points and the four regression lines.
\begin{table}[hb] 
\caption{The estimated covariance parameters.} \label{tab:var_glucose}
\small
\begin{center}
\begin{tabular}{cccc}
\hline\noalign{\smallskip}
$\sigma$ & $\sigma_a$ & $\sigma_b$ & $\rho$ \\ 
\noalign{\smallskip}\hline\noalign{\smallskip}
0.4363 & 0.4698 & 0.1156 & 0.7381 \\ 
\noalign{\smallskip}\hline
\end{tabular}
\end{center}
\end{table}
\begin{table}[hb]
\caption{The estimated random coeffiecients.} \label{tab:ran_glucose}
\small
\begin{center}
\begin{tabular}{crrrr}
\hline\noalign{\smallskip}
$i$ & blood & plasma & serum & capil   \\ 
\noalign{\smallskip}\hline\noalign{\smallskip}
$a_i$  & -0.7530 & 0.4742 & 0.2704 & 0.0084   \\
$b_i$ & -0.1007 & 0.1335 & 0.0829 & -0.1157  \\ 
\noalign{\smallskip}\hline
\end{tabular}
\end{center}
\end{table}
\noindent
\\
In Table \ref{tab:qhiGluc} we see again that everything is significant.
\begin{table}[ht]
\caption{Likelihood Ratio Test of the significance of the model terms.} \label{tab:qhiGluc}
\small
\begin{center}
\begin{tabular}{lrrr}
\hline\noalign{\smallskip}
Effect & $\chi^2$ & DF & p-value \\ 
\noalign{\smallskip}\hline\noalign{\smallskip}
Scaling, $b$  & 29.02 & 3/2 & $2.06E-07$ \\ 
Method, $a$ & 113.49 & 3/2 & $0.00E+00$    \\ 
Item, $\nu$ & 951.91 & 74.5 & $0.00E+00$ \\ 
\noalign{\smallskip}\hline
\end{tabular}
\end{center}
\end{table}

\subsubsection{Limits of Agreement}
\begin{figure}[htb!]
\centering
  \includegraphics[width=0.90\linewidth]{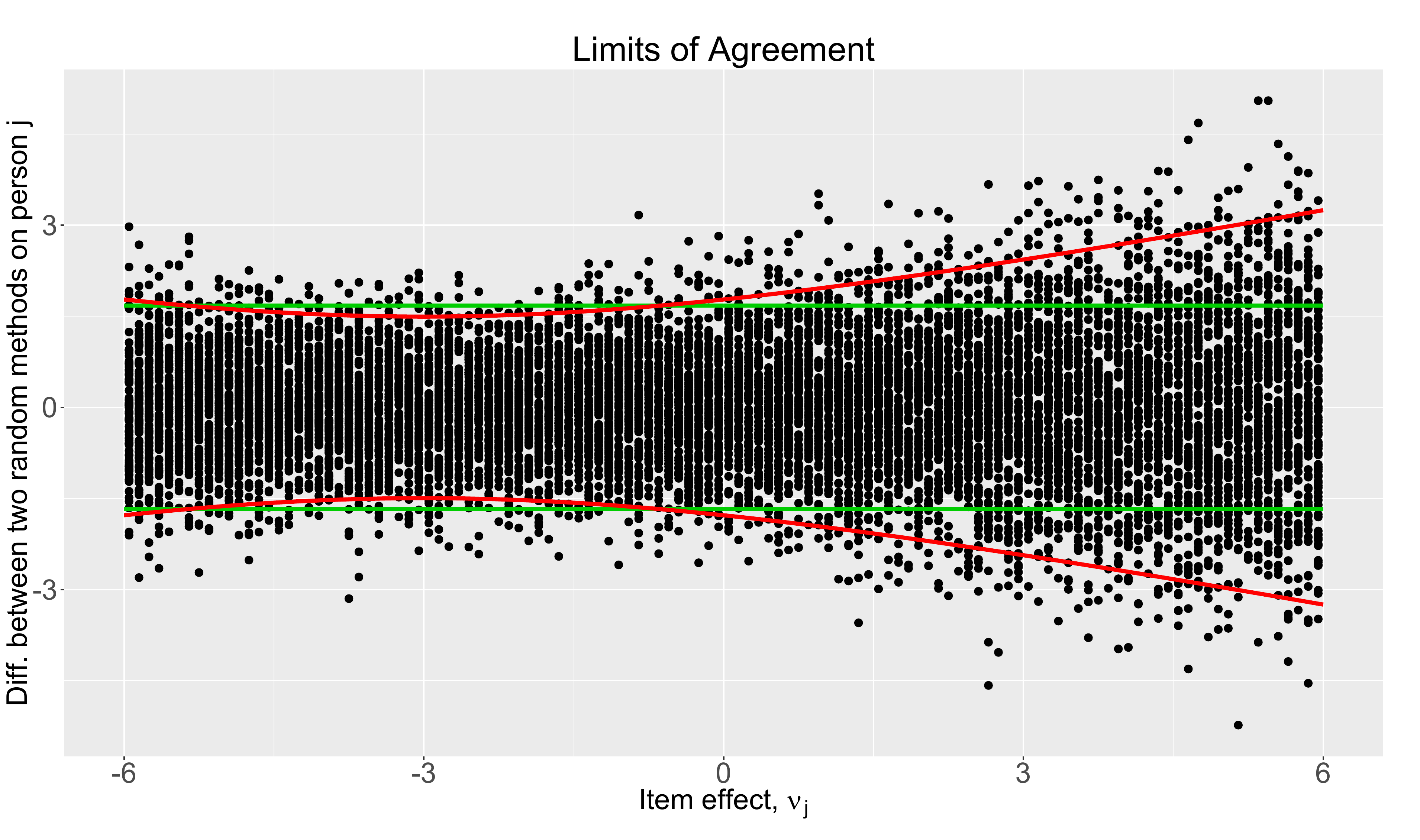}
   \caption{''Bland-Altman'' plot for simulated data and the estimated limits of agreement. We have 120 patients and 100 simulated differences for each. The green and the red lines are the limits of agreement calculated from equation (\ref{LoA2way}) and equation (\ref{LoA2multi}), respectively.} \label{fig:LoA}
\end{figure}

Agreement between methods are often assessed by estimating the limits of agreement (LoA), which give a prediction interval for the difference between measurements from two different methods on the same item. These LoAs depend on the model used and in this section it is emphasized that the intervals might become too narrow for some values of the item effect, if the multiplicative term is ignored.
\\
\\
If the simple model (\ref{2way}) is assumed to hold, the limits of agreement are
\begin{align*}
\text{LoA} = \ \ &\alpha_i - \alpha_{i'}\pm z \sqrt{\text{var}(y_{ij}-y_{i'j})} \\[2pt]= \ \ &\alpha_i - \alpha_{i'} \pm z \sqrt{\text{var}(y_{ij})+\text{var}(y_{i'j})} \\[2pt]
= \ \ &\alpha_i - \alpha_{i'} \pm z \cdot \sqrt{2\sigma^2},
\end{align*}
where $z$ is the quantile in the standard normal distribution for the desired level of the prediction interval. 
\\
\\
In \citet{ekstrom}, they consider the effect of method as random, as mentioned above. Hence they consider a mixed two-way ANOVA model, and they thus get the following limits of agreement, which is now the prediction interval for the difference between measurements from two randomly chosen methods on a new item
\begin{align} \label{LoA2way}
\text{LoA} = \ \ &0\pm z \sqrt{\text{var}(y_{ij}-y_{i'j})} \nonumber \\[2pt]
= \ \ & 0\pm z \sqrt{\text{var}(y_{ij})+\text{var}(y_{i'j})} \nonumber \\[2pt]
= \ \ &0 \pm z \cdot \sqrt{2 \cdot (\sigma_a^2+\sigma^2)}
\end{align}
This results in a wider prediction interval, since the variation between the random methods is taken into account.  
\\
\\
If the multiplicative mixed model is assumed to hold, the limits of agreement for the difference between two random methods on a new item is:
\begin{align}  \label{LoA2multi}
\text{LoA} = \ \ &0\pm z \sqrt{\text{var}(y_{ij}-y_{i'j})} \nonumber \\[2pt]
= \ \ &0\pm z \sqrt{\text{var}(y_{ij})+\text{var}(y_{i'j})} \nonumber \\[2pt]
= \ \ &0 \pm 1.96 \cdot \sqrt{2 \cdot (\sigma_a^2+\nu_j^2 \sigma_b^2+ 2 \, \nu_j \rho \, \sigma_a \sigma_b + \sigma^2)}, 
\end{align}
which gives us "trumpet shaped" prediction intervals.  
\\
\\
\textbf{Simulation Study} \\
The difference between the LoA in (\ref{LoA2way}) and  (\ref{LoA2multi}) is illustrated by a small simulation study, which simulates the following scenario: For each of 120 patients two random methods are used to measure the glucose concentration in their blood. The measurements are simulated according to the multiplicative mixed model, where the variance components are set equal to the estimates from Example 2. The method specific random effects are drawn from a multivariate normal distribution by the function $\mathsf{mvrnorm}$ from R package $\mathsf{MASS}$ \citep{RMASS}. The patients' ''true'' concentration ranges from 0.1 to 12.
\\
\\
Figure \ref{fig:LoA} shows the simulated data plotted together with the estimated limits of agreement, found by using (\ref{LoA2way}) and (\ref{LoA2multi}) on the data set from Example 2. It is clearly seen that the limits are too narrow when using (\ref{LoA2way}) for large item effects. It is also seen that for some values of the item effect, the limits are actually slightly narrower when using (\ref{LoA2multi}) instead of (\ref{LoA2way}).

\section{R package $\mathsf{mumm}$ vs. $\mathsf{NLMIXED}$ in SAS} \label{sec:sas}
The multiplicative mixed model can also be fitted by the $\mathsf{NLMIXED}$ procedure in SAS \citep{sas}. This might however be notably more time consuming than fitting the model by $\mathsf{mumm}$, if the model contains a lot of parameters. Table \ref{tab:compare} shows the computation time for the two methods when fitting the multiplicative mixed mode to three different data sets. It should be noted that PROC $\mathsf{NLMIXED}$ is run with the statement options: method=firo technique=nrridg. The computer used has 32 GB RAM, an Intel Core i7-4790 processor and runs under the operating system Windows 8.1 Enterprise. The version of R is 3.1.3, and the SAS version is 9.4.
\begin{table}[htb!]
\caption{Computation time for $\mathsf{mumm}$ and $\mathsf{NLMIXED}$.} \label{tab:compare}
\small
\begin{center}
\begin{threeparttable}
{\setlength{\extrarowheight}{5pt}%
\begin{tabular}{l|cccccc|rr}
\hline\noalign{\smallskip}\\[-19pt]
\multicolumn{1}{c|}{Data set} & \multicolumn{6}{c|}{Starting guess} & \multicolumn{2}{|c}{CPU time (sec.)} \\ 
  \cline{2-9}
\multicolumn{1}{c|}{($N_{fixed}$,$N_{random}$)\tnote{a}} & $\mu+\nu_j$ & $\sigma_a$ & $\sigma_b$ & $\sigma_d$ & $\sigma$ & $\rho$ & \multicolumn{1}{|c}{$\mathsf{mumm}$} & \multicolumn{1}{c}{$\mathsf{NLMIXED}$} \\
\noalign{\smallskip}\hline\noalign{\smallskip}
B\&O TV (12,8) & 0 & 1 & 1 & 1 & 1 & 0 & $\mathbf{0.50}$ & $\mathbf{0.68}$ \\
B\&O TV (12,8) & 5 & 0.1 & 1 & 0.5 & 1 & 0.2 & $\mathbf{0.38}$ & $\mathbf{0.48}$ \\
Wheat height (45,50)  & 0 & 1 & 1 & - & 1 & 0 & $\mathbf{9.39}$ & $\mathbf{-}$\tnote{c} \\
Wheat height (45,50) & 40 & 1 & 0.08 & - & 4 & 0 & $\mathbf{3.29}$ & $\mathbf{155.07}$ \\
Wheat height (45,50) & 89 & 2 & 0.05 & - & 6 & 0.3 & $\mathbf{1.41}$ & $\mathbf{47.87}$ \\
Hepatitis (51,6) & 0 & 1 & 1 & - & 1 & 0 & $\mathbf{1.34}$ & $\mathbf{64.71}$ \\
Hepatitis (51,6) & 8/4\tnote{b} & 0.1 & 0.05 & - & 0.2 & -0.5 & $\mathbf{0.83}$ & $\mathbf{12.30}$\\
\noalign{\smallskip}\hline
\end{tabular}}
 \begin{tablenotes}
 	\item[a] $N_{fixed}$, $N_{random}$ : the number of levels of the fixed effect and the random effect.
    \item[b] The starting guess is 8 for $j = 1,...,18$ and 4 for $j = 19,...,51$. 
    \item[c] SAS stops with the error "Optimization cannot be completed.".
  \end{tablenotes}
\end{threeparttable}
\end{center}
\end{table} 

It is seen that R package $\mathsf{mumm}$ has a lower computation time, and that the difference gets bigger when the number of parameters in the model increases or when the starting guess is far away from the optimum. Further it is seen that $\mathsf{NLMIXED}$ in one case did not succeed in finding the optimum; it stopped with the message "NRRIDG Optimization cannot be completed. Optimization routine cannot improve the function value." It is also worth noting that the "Wheat height" data set is fitted by a multiplicative mixed model without the interaction term, $d_{ij}$, even though replicates are present in the data. This is due to the fact that multilevel nonlinear mixed models are not accommodated by the $\mathsf{NLMIXED}$ procedure \citep{sasmanual}. Certain nested random effect structures can though be specified, meaning that model (\ref{eq:MMM}), including the random interaction term $d_{ij}$, \textit{can} be fitted by $\mathsf{NLMIXED}$ \citep{sasbook}. However, if the number of levels of the fixed effect is large, the implementation becomes unreasonably burdensome. 

\subsection{Confidence Intervals for Fixed Effect Contrasts} 
In equation (\ref{eq:MMM_var}) in Section \ref{sec:BO} it was shown that the variance of the product difference depends on the actual size of the difference itself, with increasing variance for increasing contrast. This means that proper confidence intervals for product contrasts are asymmetric; they are wider "away from zero" than "towards zero" \citep{MAM}. This result is obviously valid in the other applications as well. However, the $\mathsf{NLMIXED}$ procedure only provides symmetrical t-distribution based confidence intervals, whereas R package $\mathsf{mumm}$ makes is possible to obtain profile likelihood based confidence intervals, which will be asymmetric due to the nature of the model. 

\section{Concluding Remarks} \label{sec:conc}
In every situation where an interaction between a fixed effect and a random effect is present, improved inference might be possible by extending the usual linear mixed model with a multiplicative term. The purpose of this term is to obtain a more nuanced modelling and interpretation of the interaction, leading to a more correct inference of the model effects. 
The version of the multiplicative mixed model studied in this paper was shown to have multiple applications, and with the newly developed R package $\mathsf{mumm}$, this non-linear mixed model is easily fitted to data. 

\section{Conflict of Interest}
The authors declare that they have no conflict of interest.
\FloatBarrier
\appendix
\pagebreak
\section{R-code for B\&0 TV Data} \label{app}
\begin{lstlisting}[language=R, columns=fixed, frame=single, numbers = none, basicstyle = \ttfamily\small, caption={R-code used to produce the results in Section \ref{sec:BO}.}]
library(SensMixed); library(lmerTest); library(mumm)

# .......................... Loading The BO TV Data ..........................
DATA = TVbo
DATA$Product = as.factor(DATA$TVset:DATA$Picture)

BO = data.frame(DATA$Product)
names(BO) = "Product"
BO$Assessor = DATA$Assessor
BO$Cutting = DATA$Cutting
BO$Sharpmove = DATA$Sharpnessofmovement

# ........... Fitting the multiplicative mixed model with mumm ..............
fit = mumm(Cutting ~ 1 + Product + (1|Assessor) +
             (1|Assessor:Product) + mp(Assessor,Product), data = BO)

a = mumm::ranef(fit)$Assessor
b = mumm::ranef(fit)$`mp Assessor:Product`

# ...................... Likelihood ratio-tests .............................
fit_no_d =  mumm(Cutting ~ 1 + Product + (1|Assessor) +
                  mp(Assessor,Product),data = BO)
fit_no_a =  mumm(Cutting ~ 1 + Product + (1|Assessor:Product) +
                  mp(Assessor,Product),data = BO)
fit_no_b =  lmer(Cutting ~ 1 + Product + (1|Assessor) +
                  (1|Assessor:Product),data = BO, REML = FALSE)
fit_no_v =  lmer(Cutting ~ 1 + (1|Assessor) + (1|Assessor:Product),
                data = BO, REML = FALSE)

#The chi^2 values:
X2 = 2*c(-fit$objective-(-fit_no_d$objective), # Test of disagreement.
         -fit$objective-logLik(fit_no_b),      # Test of scaling.
         -fit$objective-(-fit_no_a$objective), # Test of assessor.
         -fit$objective-logLik(fit_no_v)       # Test of product.
         );
pvalues = 1-pchisq(X2, df = c(1/2,3/2,3/2,12.5));

# ....................... Detecting product differences........................
# Linear 2-way mixed model:
fit_2way = lmer(Sharpmove ~ 1 + Product  + (1|Assessor) +
                  (1|Assessor:Product),data = BO, REML = TRUE)
anova(fit_2way)

# Mixed Assessor Model:
fit_MAM = sensmixed(c("Sharpmove","Sharpmove"), 
                    c("Product"),c("Assessor"),
                    data = BO, product_structure = 1, 
                    error_structure = "ONLY-ASS", MAM = TRUE,
                    control = sensmixedControl(calc_post_hoc = TRUE, 
                                               MAM_balanced = TRUE,
                                               MAM_adjusted = FALSE))
fit_MAM[[3]]

# Multiplicative mixed model:
fit2 = mumm(Sharpmove ~ 1 + Product + (1|Assessor) +
            (1|Assessor:Product) + mp(Assessor,Product), data = BO)
fit2_nov = lmer(Sharpmove ~ 1 + (1|Assessor) + (1|Assessor:Product),
                data = BO, REML = FALSE)
X2 = 2*(-fit2$objective-logLik(fit2_nov))
1-pchisq(X2, df = 12.5);
\end{lstlisting} 
\FloatBarrier

\bibliographystyle{plainnat}
\bibliography{references}

\end{document}